\documentclass[twoside,twocolumn,english,pra,show pacs]{revtex4}
\usepackage[T1]{fontenc}
\usepackage[latin9]{inputenc}
\usepackage{babel}
\usepackage{amstext}
\usepackage{amssymb}
\usepackage{graphicx}
\usepackage{esint}
\usepackage[unicode=true,pdfusetitle,
 bookmarks=true,bookmarksnumbered=false,bookmarksopen=false,
 breaklinks=false,pdfborder={0 0 1},backref=false,colorlinks=false]
 {hyperref}
\usepackage{breakurl}

\makeatletter
\@ifundefined{textcolor}{}
{%
 \definecolor{BLACK}{gray}{0}
 \definecolor{WHITE}{gray}{1}
 \definecolor{RED}{rgb}{1,0,0}
 \definecolor{GREEN}{rgb}{0,1,0}
 \definecolor{BLUE}{rgb}{0,0,1}
 \definecolor{CYAN}{cmyk}{1,0,0,0}
 \definecolor{MAGENTA}{cmyk}{0,1,0,0}
 \definecolor{YELLOW}{cmyk}{0,0,1,0}
}

\usepackage{multirow}

\makeatother

\begin{document}

\title{Tuning linear response dynamics near the Dirac points\\in the bosonic
Mott insulator}

\author{A. S. Sajna}
\begin{abstract}
Optical lattice systems offer the possibility of creating and tuning
Dirac points which are present in the tight-binding lattice dispersions.
For example, such a behavior can be achieved in the staggered flux
lattice or honeycomb type of lattices. Here we focus on the strongly
correlated bosonic dynamics in the vicinity of Dirac points. In particular,
we investigate bosonic Mott insulator phase in which quasiparticle
excitations have a simple particle-hole interpretation. We show that
linear response dynamics around Dirac points, can be significantly
engineered at least in two ways: by the type of external perturbation
or by changing the lattice properties. The key role is played by the
interband transitions. Moreover, we explain that the behavior of these
transitions is directly connected to different energy scales of the
effective hopping amplitudes for particles and holes. Presented in
this work theoretical study about tunability of linear response dynamics
near the Dirac points can be directly simulated in the optical lattice
systems.
\end{abstract}


\address{Faculty of Physics, Adam Mickiewicz University, ul. Umultowska 85,
61-614 Pozna\'{n}, Poland}

\maketitle

\section{Introduction}

Response of the ultracold atomic systems to the external periodic
modulation has been widely investigated experimentally and theoretically
within the context of the Bose Hubbard model (BHM) (see, e.g., \cite{PhysRevLett.92.130403,PhysRevLett.93.240402,PhysRevA.72.063609,PhysRevLett.97.050402,PhysRevA.73.041608,Huber:2007uq,2011PhRvL.106t5301T,amplitude-modulation-Hou,PhysRevLett.107.175301,2012Natur.487..454E,Lacki-Zakrzewski-PhysRevA.86.013602,PhysRevLett.109.010401,PhysRevA.89.063623,Strand2015,PhysRevB.92.174521,PhysRevA.94.043612}).
In particular, periodic modulation protocols help in understanding
of strongly correlated bosonic dynamics, e.g. its gapped nature in
the bosonic Mott insulator (MI) \cite{PhysRevLett.92.130403}, dynamical
conductivity \cite{2011PhRvL.106t5301T}, collective Higgs modes \cite{2012Natur.487..454E}
or thermal excitations \cite{PhysRevA.72.063609,PhysRevA.94.043612}.
However so far, these problems are poorly understood within the context
of non-trivial lattices which can be generated by optical lattice
patterns, e.g. geometrically or by synthetic gauge potentials (\cite{Goldman2014,2012Natur.483..302T,Windpassinger:2013uf}
and literature therein). This non-triviality in dynamics can emerge
for example from Dirac points which can be present in the tight binding
energy dispersion \cite{2012Natur.483..302T}. Such a problem is especially
interesting because the resulting dynamics exhibits much broader types
of quasiparticle excitations, e.g. intra and interband transitions.
The lattices that show such a dynamics have been very recently studied
by Grygiel et al. \cite{PhysRevB.96.094520}. They have investigated
optical conductivity in the lattice with synthetic gauge potential
which correspond to the uniform magnetic field. The Grygiel et al.
work has extended the earlier one \cite{2014PhRvA..89b3631S,PhysRevA.90.043603,Sajna2015acta,Grygiel2016}
and in particular they have shown the importance of interband transitions
around Dirac points for two-band system with uniform $\pi$ flux.

In this work, we present that non-trivial dynamics around Dirac points
also appears in the broad class of lattices with staggered symmetry.
In particular, in comparison to Ref. \cite{PhysRevB.96.094520}, interband
transitions can be also very clearly visible in a much simpler experimentally
available optical lattice systems without gauge potential, i.e. in
the honeycomb type of lattices \cite{2012Natur.483..302T}.

As the main point of our studies we show by using analytical and numerical
arguments that such interband transitions are very sensitive to the
type of experiment made. In particular, we show that these transitions
can be engineered at least in two ways: by a suitable choice of external
perturbation or by changing the lattice properties. In the first case
we show that the interband transitions can be simply turned off in
the isotropic amplitude modulation of the lattice \cite{Huber:2007uq,PhysRevLett.109.010401,2012Natur.487..454E}
in contrary to the phase modulation protocols \cite{2011PhRvL.106t5301T}.
In the latter case, interband transitions can be continuously modified
by engineering of Dirac like physics in staggered symmetry lattices
\cite{2010PhRvA..81b3404L,2008PhRvL.100m0402L,2012Natur.483..302T,PhysRevLett.110.165304}.
As an example, we analyze honeycomb type lattices which are able to
shift the location of Dirac cones within the Brillouin zone \cite{2012Natur.483..302T}.
We also consider the lattice with staggered fluxes for which a tunability
of relativistic dynamics can be broadly modified by the steepness
of Dirac cones simply observed in the tight-binding dispersion energy
\cite{2008PhRvL.100m0402L,2010PhRvA..81b3404L}. Therefore, we show
that engineering of Dirac like physics significantly modifies the
linear response dynamics and we explain that this allows changes in
the energy absorption rate mostly locally around Dirac points.

Through the paper we focus on the bosonic Mott insulator for which
quasiparticle excitations have simple particle-hole interpretation
and all peculiar dynamical phenomena appear at frequencies proportional
to the bosonic on-site interaction \cite{2014PhRvA..89b3631S,PhysRevA.90.043603}.

Manuscript is organized as follows. In Sec. \ref{sub: Model-and-effective}
and \ref{sub:Staggered-symmetry-lattice } we shortly introduce the
BHM within the coherent path integral framework and we define lattices
types which are analyzed in this paper. Next, in Sec. \ref{sub: Linear-response-kernel},
\ref{sub:Dynamical-conductivity} and \ref{sub: Isotropic-energy-absorption},
we discuss the linear response dynamics together with their application
to the conductivity and the isotropic energy absorption rate. In Sec.
\ref{sec: Summary} we summarize our results.

\section{Methods and results \label{sec:Model}}

\subsection{Model and effective action \label{sub: Model-and-effective}}

We consider strongly correlated lattice bosons which are described
by the Bose-Hubbard model \cite{Fisher:1989zza,1963PhRv..129..959G}.
Hamiltonian of this model in the second quantization language has
a form
\begin{equation}
H=H_{0}+H_{1},\label{eq:BHM-hamiltonian}
\end{equation}
\begin{equation}
H_{0}=\frac{U}{2}\sum_{i}\hat{b}_{i}^{\dagger}\hat{b}_{i}^{\dagger}\hat{b}_{i}\hat{b}_{i}-\mu\sum_{i}\hat{b}_{i}^{\dagger}\hat{b}_{i}\,,\label{eq:BHM-H0}
\end{equation}
\begin{equation}
H_{1}=-J\sum_{\langle ij\rangle}\hat{b}_{i}^{\dagger}\hat{b}_{j}\,,\label{eq: H1}
\end{equation}
where $\hat{b}_{i}$ ($\hat{b}_{i}^{\dagger}$) is the annihilation
(creation) bosonic operator at a site $i$. The parameters $J$, $U$
and $\mu$ correspond to the hopping, interaction and chemical potential
energy, respectively. We assume that the sum in Eq. (\ref{eq: H1})
is restricted to the nearest neighbours sites. 

Using the coherent state path integral formalism we can obtain the
following form of partition function \cite{Stoof:2009tf}
\begin{equation}
\mathcal{Z}=\int\mathcal{D}b^{*}\mathcal{D}b\, e^{-\left(S_{0}+S_{1}\right)/\hbar},\label{eq:statistical-sum-0}
\end{equation}
\begin{eqnarray}
S_{0} & = & \sum_{i}\int_{_{0}}^{\hbar\beta}d\tau\left\{ \bar{b}_{i}(\tau)\hbar\partial_{\tau}b_{i}(\tau)\right.\label{eq:actionprzed1}\\
 &  & \left.+\frac{U}{2}\bar{b}_{i}(\tau)\bar{b}_{i}(\tau)b_{i}(\tau)b_{i}(\tau)-\mu b_{i}^{*}(\tau)b_{i}(\tau)\right\} ,\nonumber 
\end{eqnarray}
\begin{equation}
S_{1}=-J\sum_{\langle ij\rangle}\int_{_{0}}^{\hbar\beta}d\tau\;\bar{b}_{i}(\tau)b_{j}(\tau),\label{eq:actionprzed3}
\end{equation}
In this work, we are interested in the bosonic MI phase which appears
for $J\ll U$ and for integer densities. Therefore we treat $S_{1}$
term in Eq. (\ref{eq:actionprzed3}) as a perturbation. One of the
powerful methods in this regime is based on the strong coupling approach
proposed by Sengupta and Dupuis \cite{2005PhRvA..71c3629S}. This
method uses double Hubbard-Stratonovich (HS) transformations which
allow the identification of new HS fields with that from Eqs. (\ref{eq:actionprzed1})-(\ref{eq:actionprzed3}).
Such an identification is possible because of correlation functions
correspondence between the bare fields $b_{i}(\tau)$, $\bar{b}_{i}(\tau)$
from Eqs. (\ref{eq:actionprzed1})-(\ref{eq:actionprzed3}) and the
fields after the second HS \cite{2005PhRvA..71c3629S}. Then, the
effective action takes the form 
\begin{eqnarray}
 &  & S_{eff}=-\sum_{ij}\int_{0}^{\beta}d\tau J_{ij}\bar{b}_{i}(\tau)b_{j}(\tau)\nonumber \\
 &  & -\sum_{i}\int_{0}^{\beta}d\tau d\tau'\left[G^{1,c}\left(\tau'-\tau\right)\right]^{-1}\bar{b}_{i}(\tau')b_{i}(\tau),\label{eq:Seff}
\end{eqnarray}
where the strong coupling expansion is truncated to the second order
in the MI phase in which Gaussian fluctuations make a good approximation
\cite{Stoof:2009tf}. Moreover, $\beta$ is the inverse of temperature
i.e. $1/k_{B}T$ ($k_{B}$ is the Boltzmann constant). Starting from
Eq. (\ref{eq:Seff}) we set the reduced Planck constant $\hbar$ to
unity for simplicity. Moreover $G^{1,c}\left(\tau-\tau'\right)$ is
the two-point local Green function which can be defined by its Fourier
transform in the Matsubara frequencies domain $\omega_{n}$, i.e.
$G^{1,c}\left(\tau\right)=\frac{1}{\beta}\sum_{\omega_{n}}G^{1,c}\left(\omega_{n}\right)e^{i\omega_{n}\tau}$
where
\begin{equation}
G^{1,c}\left(i\omega_{n}\right)=\frac{n_{0}+1}{i\omega_{n}+E_{n_{0}}-E_{n_{0}+1}}-\frac{n_{0}}{i\omega_{n}+E_{n_{0}-1}-E_{n_{0}}},
\end{equation}
with the on-site energy $E_{n_{0}}=-\mu n_{0}+Un_{0}(n_{0}-1)/2$
and $G^{1,c}\left(i\omega_{n}\right)$ is taken in the zero temperature
limit. The Matsubara frequencies are defined as $\omega_{n}=2\pi n/\beta$
and $n$ is an integer number.

\subsection{Lattices with staggered symmetry \label{sub:Staggered-symmetry-lattice }}

In this work we consider two types of lattices whose tight-binding
energy dispersions exhibit relativistic behavior. In the first type
of lattices, the relativistic energy behavior is introduced by the
synthetic gauge field i.e. staggered flux lattice \cite{2008PhRvL.100m0402L,2010PhRvA..81b3404L}
and in the second one by the lattice geometry: honeycomb or brick-wall
lattice \cite{2012Natur.483..302T,PhysRevLett.110.165304}. Then,
the effective action from Eq. (\ref{eq:Seff}) in the wave vector
basis $\mathbf{k}=\left(k_{x},\, k_{y}\right)$, can be rewritten
to the form
\begin{eqnarray}
S^{eff} & = & -\sum_{\mathbf{k}}\int d\tau d\tau'\mathcal{B}_{\mathbf{k}}^{\dagger}(\tau')\left[G^{MI}(\mathbf{k},\tau'-\tau)\right]^{-1}\mathcal{B}_{\mathbf{k}}(\tau),\label{eq: effective action}
\end{eqnarray}
where
\begin{equation}
\left[G^{MI}(\mathbf{k},\tau'-\tau)\right]^{-1}=-\mathbf{F}(\mathbf{k})\delta\left(\tau'-\tau\right)+G_{0}^{-1}\left(\tau'-\tau\right),\label{eq: MI Green funtion in tau}
\end{equation}
\begin{equation}
\mathbf{F}(\mathbf{k})=\left[\begin{array}{cc}
0 & f_{1,2}(\mathbf{k})\\
\bar{f}_{1,2}(\mathbf{k}) & 0
\end{array}\right],\label{eq: general TB}
\end{equation}
\begin{equation}
\mathcal{B}_{\mathbf{k}}(\tau)=\left[\begin{array}{c}
b_{\mathbf{k}A}(\tau)\\
b_{\mathbf{k}B}(\tau)
\end{array}\right],
\end{equation}
and where indices $A$ and $B$ label the fields which belong to the
corresponding sublattices and moreover we also assume that $\bar{f}_{1,2}(\mathbf{k})=f_{2,1}(\mathbf{k})$.
In the subsequent calculations, we consider three forms of $f_{1,2}(\mathbf{k})$
which correspond to the three different lattices exhibiting Dirac
points.

- Staggered flux lattice
\begin{equation}
f_{1,2}^{flux}(\mathbf{k})=-2J\left[e^{i\phi/4}\cos\left(k_{x}a\right)+e^{-i\phi/4}\cos\left(k_{y}a\right)\right],\label{eq: lattice flux}
\end{equation}
where $\phi$ is the amplitude of the staggered flux \cite{2008PhRvL.100m0402L,2010PhRvA..81b3404L}.

- Honeycomb lattice 
\begin{equation}
f_{1,2}^{hc}(\mathbf{k})=-J\left(e^{-ik_{y}a}+e^{i\frac{1}{2}k_{y}a-i\frac{\sqrt{3}}{2}k_{x}a}+e^{i\frac{1}{2}k_{y}a+i\frac{\sqrt{3}}{2}k_{x}a}\right).\label{eq: lattice honeycomb}
\end{equation}

- Brick-wall lattice 
\begin{equation}
f_{1,2}^{bw}(\mathbf{k})=-J\left(e^{ik_{x}a}+e^{-ik_{x}a}+e^{-k_{y}a}\right).\label{eq: lattice brick-wall}
\end{equation}
In the above equations, $a$ is the lattice constant.

\begin{figure}[th]
\includegraphics[scale=0.67]{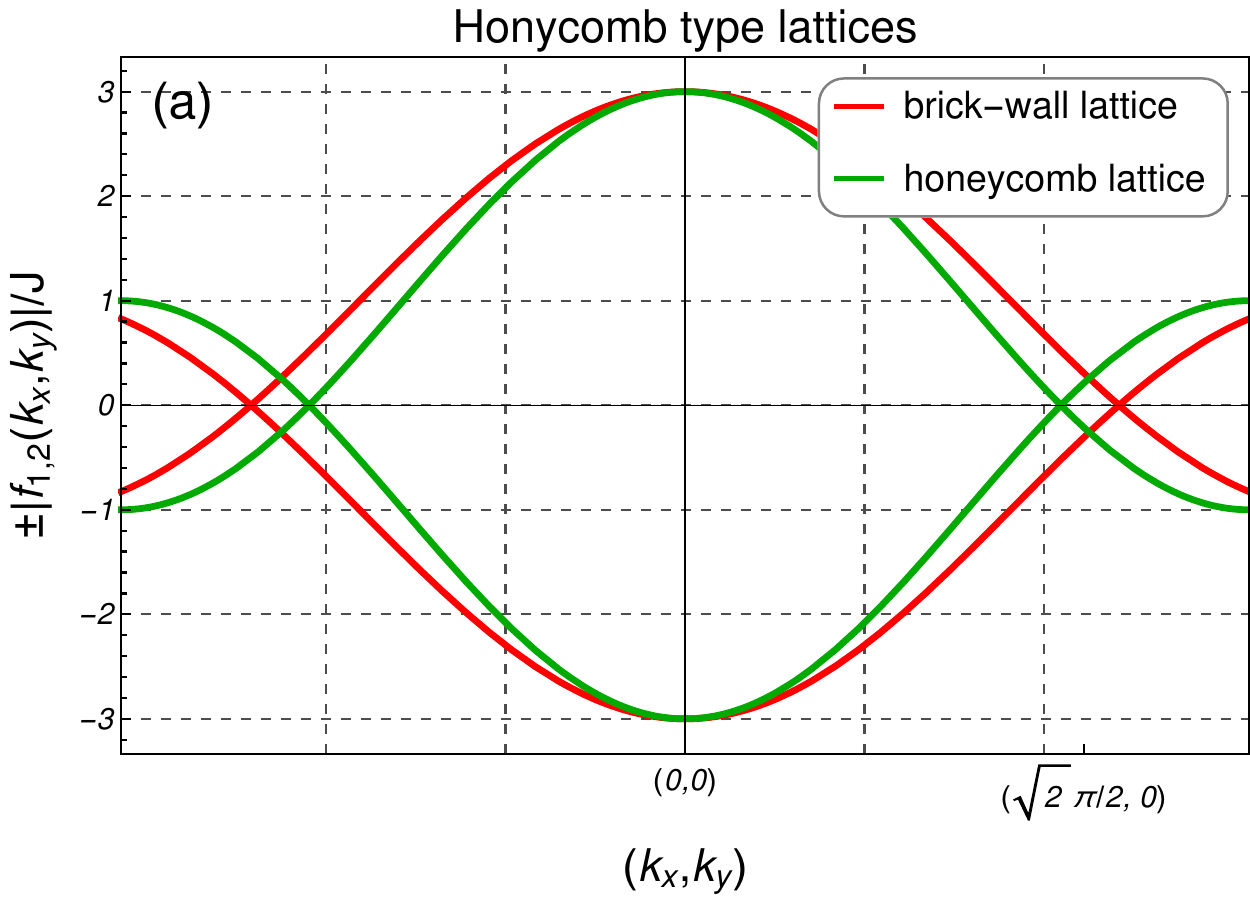}

\includegraphics[scale=0.67]{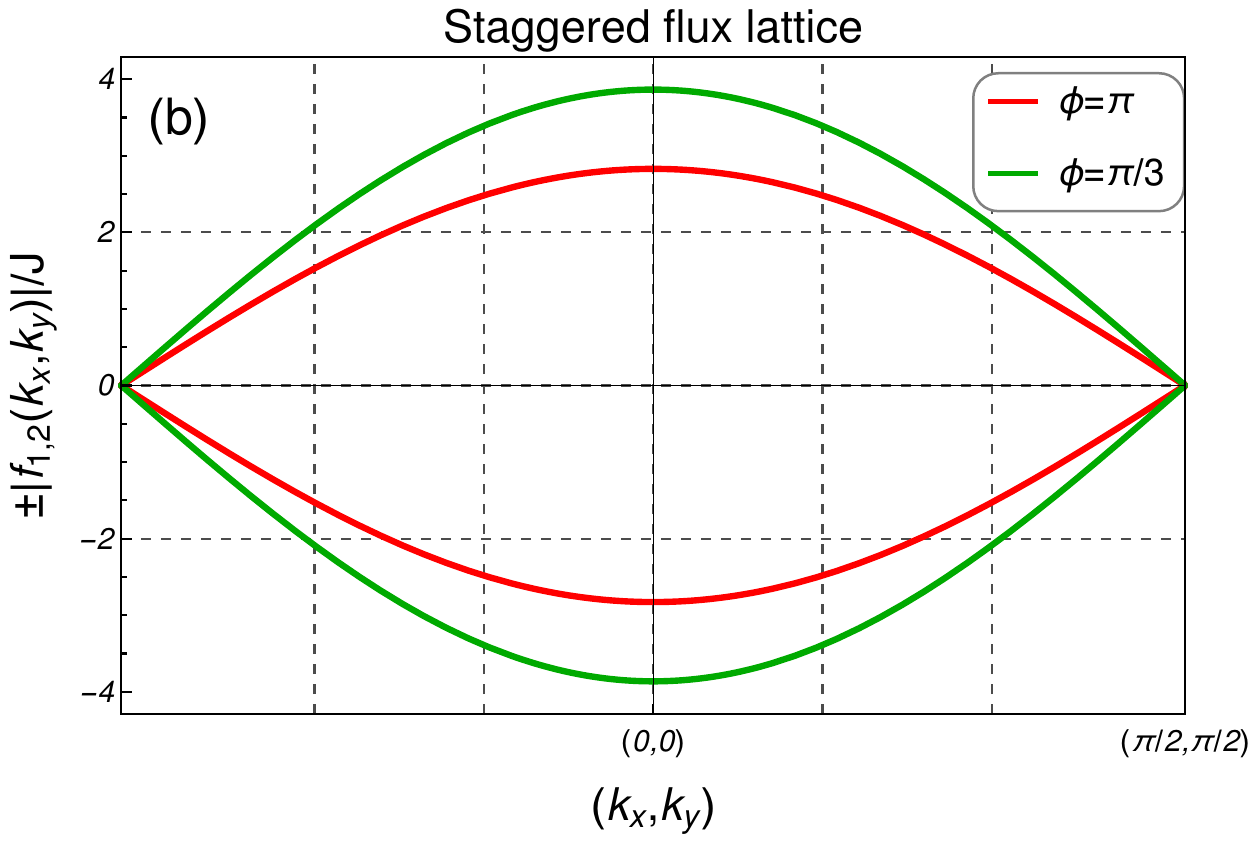}\caption{(color online) The tight-binding energy dispersion for (a) the honeycomb
type of lattices - standard honeycomb and brick-wall lattices, (b)
the staggered flux lattice with $\phi=\pi$ and $\pi/3$. The particular
paths in the wave vector space are chosen to show the existence of
Dirac points in the tight-binding dispersion, i.e. the points at which
dispersions intersects.\label{fig: tightbinding and diracs}}
\end{figure}

\begin{figure}[th]
\includegraphics[scale=0.67]{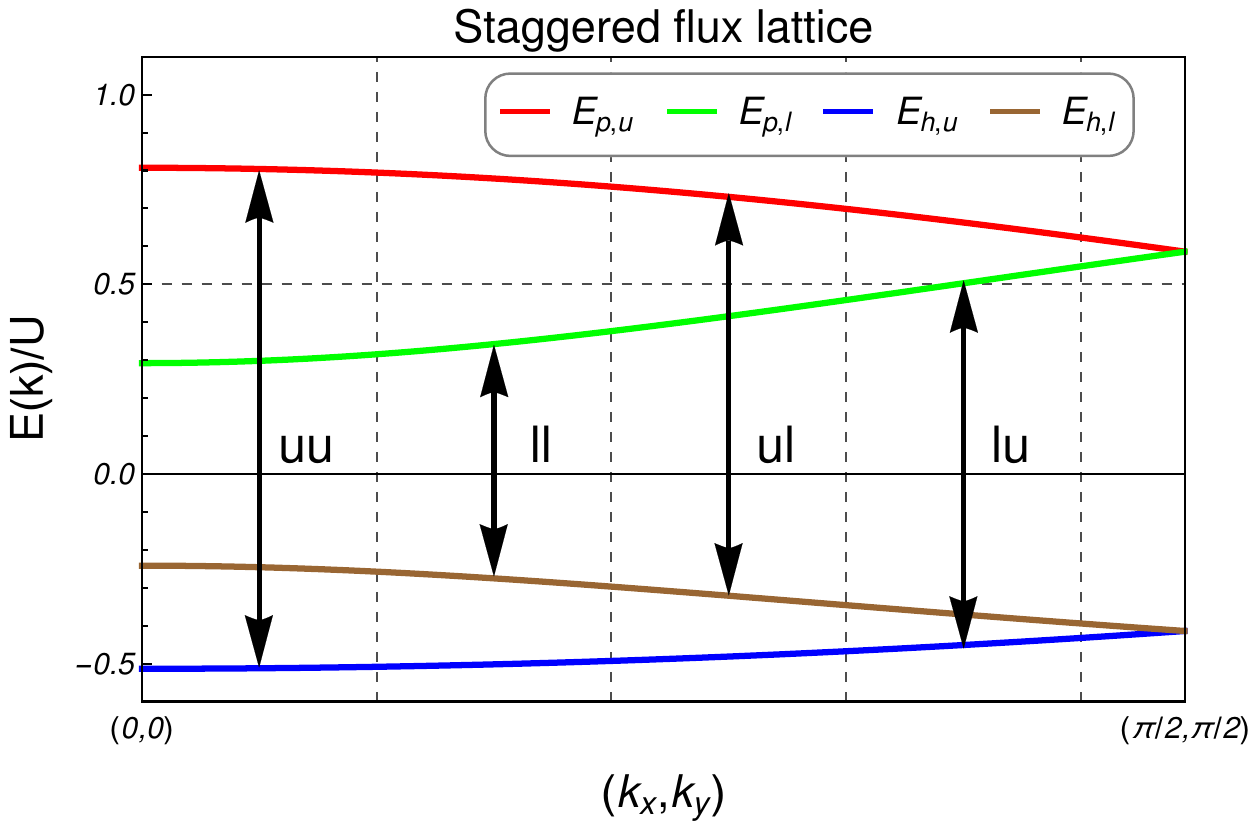}\caption{(color online) The wave vector $\mathbf{k}=\left(k_{x},\, k_{y}\right)$
dependence of the quasi-particle and hole energy bands in the MI phase.
The staggered flux lattice for $J/U=0.043$ and $\mu=0.414$ is depicted.
The arrows show possible transitions in the linear response dynamics
considered in this paper. $uu$, $ll$ and $ul$, $lu$ correspond
to the intra and inter-band transitions, respectively. \label{fig: TRANSITIONS}}
\end{figure}

At the end of this section we also give the form of MI Green function
$G^{MI}$ (see Eq. (\ref{eq: MI Green funtion in tau})) in the Matsubara
frequency representation $\omega_{n}$ which will be useful in further
calculations, i.e.
\begin{equation}
G^{MI}(\mathbf{k},i\omega_{n})=\left[\begin{array}{cc}
G_{1,1}^{MI}(\mathbf{k},\omega_{n}) & G_{1,2}^{MI}(\mathbf{k},\omega_{n})\\
G_{2,1}^{MI}(\mathbf{k},\omega_{n}) & G_{2,2}^{MI}(\mathbf{k},\omega_{n})
\end{array}\right],
\end{equation}
where diagonal terms $\alpha=\beta$ are
\begin{eqnarray}
 &  & G_{\alpha,\beta}^{MI}(\mathbf{k},\omega_{n})\nonumber \\
 &  & =\frac{1}{2}\left[\frac{z_{p,u}\left(\mathbf{k}\right)}{i\omega_{n}-E_{p,u}\left(\mathbf{k}\right)}+\frac{1-z_{p,u}\left(\mathbf{k}\right)}{i\omega_{n}-E_{h,u}\left(\mathbf{k}\right)}\right.\nonumber \\
 &  & \left.-\frac{z_{p,l}\left(\mathbf{k}\right)}{i\omega_{n}-E_{p,l}\left(\mathbf{k}\right)}-\frac{1-z_{p,l}\left(\mathbf{k}\right)}{i\omega_{n}-E_{h,l}\left(\mathbf{k}\right)}\right],\label{eq: Green function diagonal}
\end{eqnarray}
and off-diagonal terms $\alpha\neq\beta$ are
\begin{eqnarray}
 &  & G_{\alpha,\beta}^{MI}(\mathbf{k},\omega_{n})=\frac{f_{\alpha,\beta}(\mathbf{k})}{2\left|f_{\alpha,\beta}(\mathbf{k})\right|}\nonumber \\
 &  & \times\left[-\frac{z_{p,u}\left(\mathbf{k}\right)}{i\omega_{n}-E_{p,u}\left(\mathbf{k}\right)}-\frac{1-z_{p,u}\left(\mathbf{k}\right)}{i\omega_{n}-E_{h,u}\left(\mathbf{k}\right)}\right.\nonumber \\
 &  & \left.+\frac{z_{p,l}\left(\mathbf{k}\right)}{i\omega_{n}-E_{p,l}\left(\mathbf{k}\right)}+\frac{1-z_{p,l}\left(\mathbf{k}\right)}{i\omega_{n}-E_{h,l}\left(\mathbf{k}\right)}\right].\label{eq: Green function off-diagonal}
\end{eqnarray}
The corresponding $z_{p,\alpha}\left(\mathbf{k}\right)$ and $E_{p,u/l}\left(\mathbf{k}\right)$
quantities in Eqs. (\ref{eq: Green function diagonal})-(\ref{eq: Green function off-diagonal})
are
\begin{equation}
z_{p,\alpha}\left(\mathbf{k}\right)=\frac{E_{p,\alpha}\left(\mathbf{k}\right)+\mu+U}{E_{p,\alpha}\left(\mathbf{k}\right)-E_{h,\alpha}\left(\mathbf{k}\right)}\;,
\end{equation}
\begin{equation}
E_{p,u/l}\left(\mathbf{k}\right)=\frac{\pm\left|f_{1,2}(\mathbf{k})\right|}{2}-\mu+U\left(n_{0}-\frac{1}{2}\right)\pm\frac{1}{2}\Delta_{u/d}\left(\mathbf{k}\right),
\end{equation}
\begin{equation}
\Delta_{u/l}\left(\mathbf{k}\right)=\sqrt{\left|f_{1,2}(\mathbf{k})\right|^{2}\pm4\left|f_{1,2}(\mathbf{k})\right|U\left(n_{0}+\frac{1}{2}\right)+U^{2}}.\label{eq:gap}
\end{equation}
$E_{p,u/l}\left(\mathbf{k}\right)$ and $E_{h,u/l}\left(\mathbf{k}\right)$
are poles of the Green functions in Eqs. (\ref{eq: Green function diagonal})-(\ref{eq: Green function off-diagonal}),
therefore they are interpreted as quasiparticle energies. The $p$
($h$) index corresponds to the quasiparticle (quasihole) excitation,
while $u$ ($l$) corresponds to the upper (lower) band in the tight-binding
model. The $u$ and $l$ bands in the tight binding model are obtained
from the eigenenergies of $\mathbf{F}(\mathbf{k})$ (Eq. (\ref{eq: general TB}))
and are given by $\left|f_{1,2}(\mathbf{k})\right|$ and $-\left|f_{1,2}(\mathbf{k})\right|$,
respectively. 

An important feature of the lattices considered here is that they
exhibit tunable Dirac points. To show this, we plot the tight-binding
energy dispersions $\pm\left|f_{1,2}(\mathbf{k})\right|$ of honeycomb
and brick-wall lattices in Fig. \ref{fig: tightbinding and diracs}
a and the tight-binding energy dispersions of staggered flux lattice
in Fig. \ref{fig: tightbinding and diracs} b. The corresponding formulas
for these dispersions are given in Appendix \ref{sub:Tight-binding-energy-dispersions}.
The tunability of the honeycomb like lattice is obtained by modification
of its geometry (standard honeycomb geometry and brick-wall geometry)
which cause a shift of Dirac points locations in the wave vector space
(see Fig. \ref{fig: tightbinding and diracs} a). This modification
also slightly changes the steepness of the relativistic dispersion
around Dirac points. Moreover, changing the flux amplitude $\phi$
in the staggered flux lattice implies changes in the steepness of
relativistic dispersion around the Dirac points, which is more pronounced
than in Fig. \ref{fig: tightbinding and diracs} a. These effects
have direct consequences on the linear response dynamics which will
be discussed in the subsequent sections. 

Additionally, to show how the excitation spectra of quasiparticles
are affected by the Dirac points, we plot the MI energy dispersion
of staggered flux lattice in Fig. \ref{fig: TRANSITIONS} (the analogous
plot can be given for the honeycomb like lattices). In this Figure,
the Dirac points are located in the quasiparticle and hole spectrum
at $\left(\pi/2,\,\pi/2\right)$ and are separated by energy $U$
(a similar observation was made in the uniform magnetic field \cite{2014PhRvA..89b3631S,PhysRevA.90.043603}).

\begin{figure*}[th]
\includegraphics[scale=0.47]{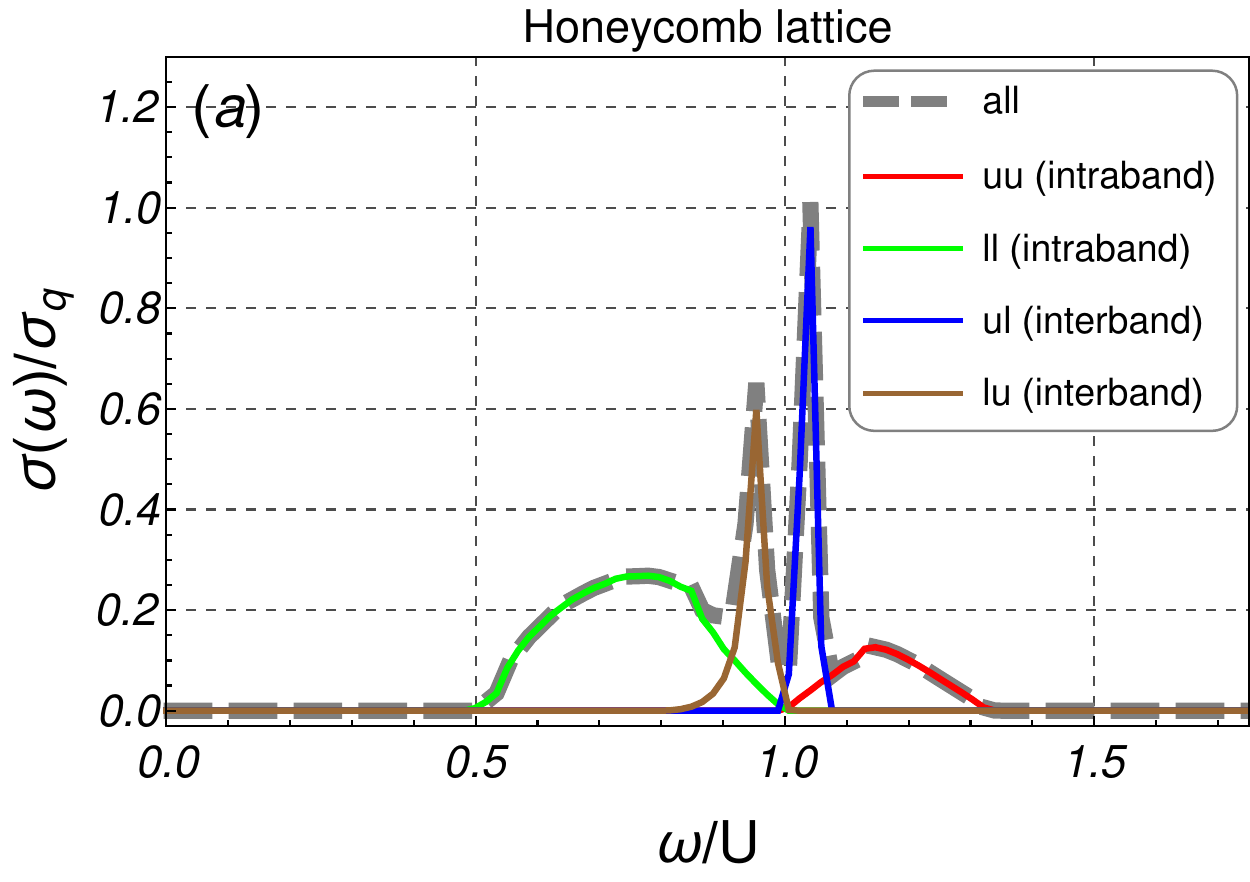}\includegraphics[scale=0.47]{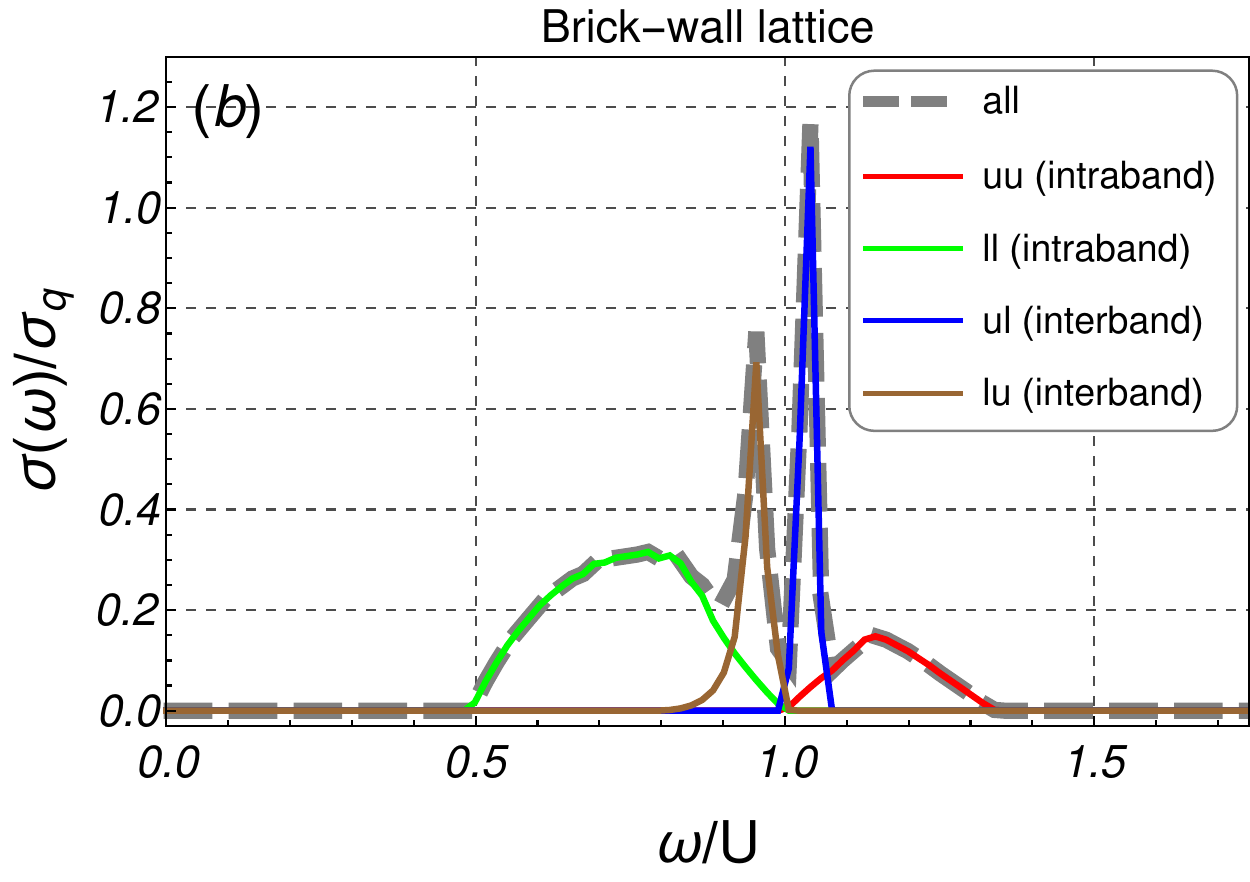}\includegraphics[scale=0.47]{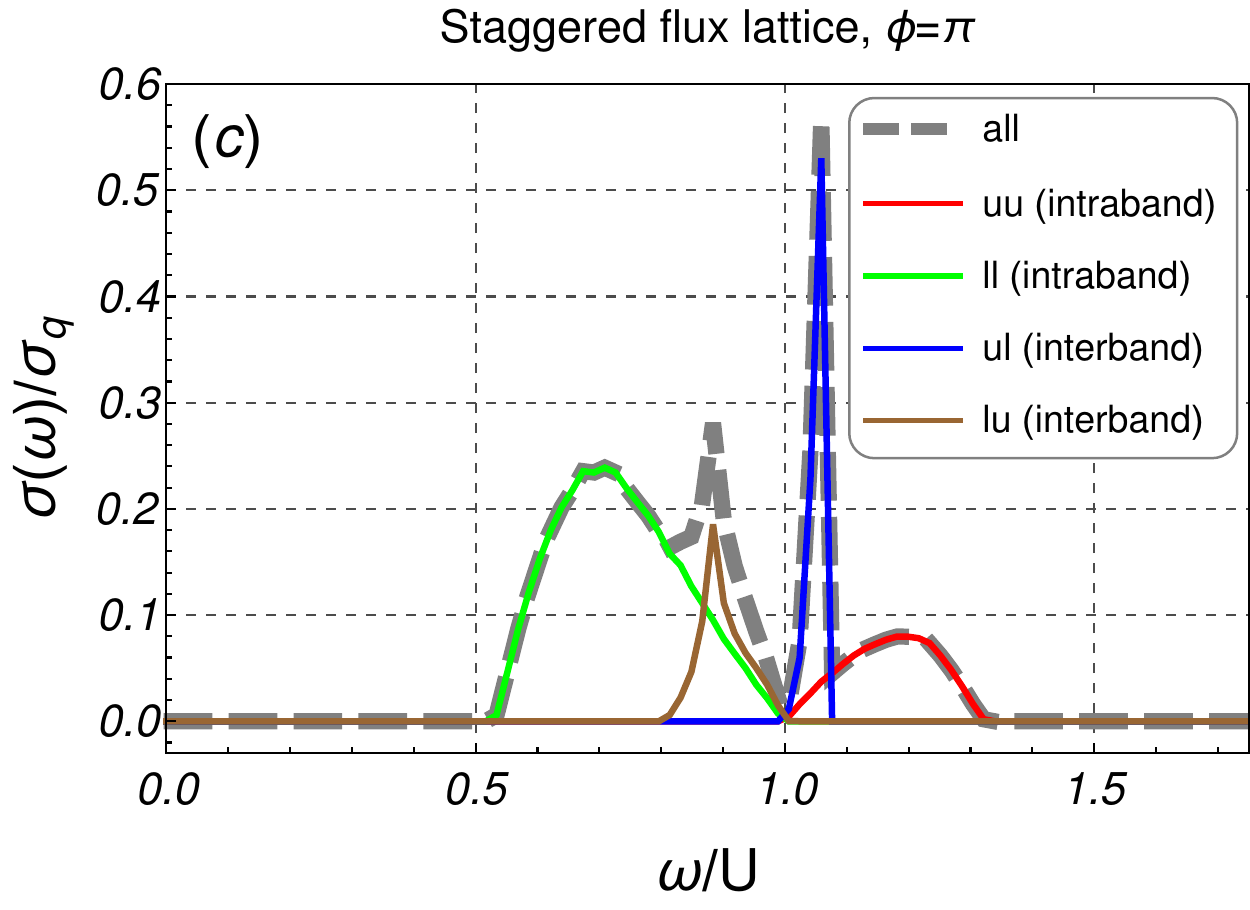}

\includegraphics[scale=0.47]{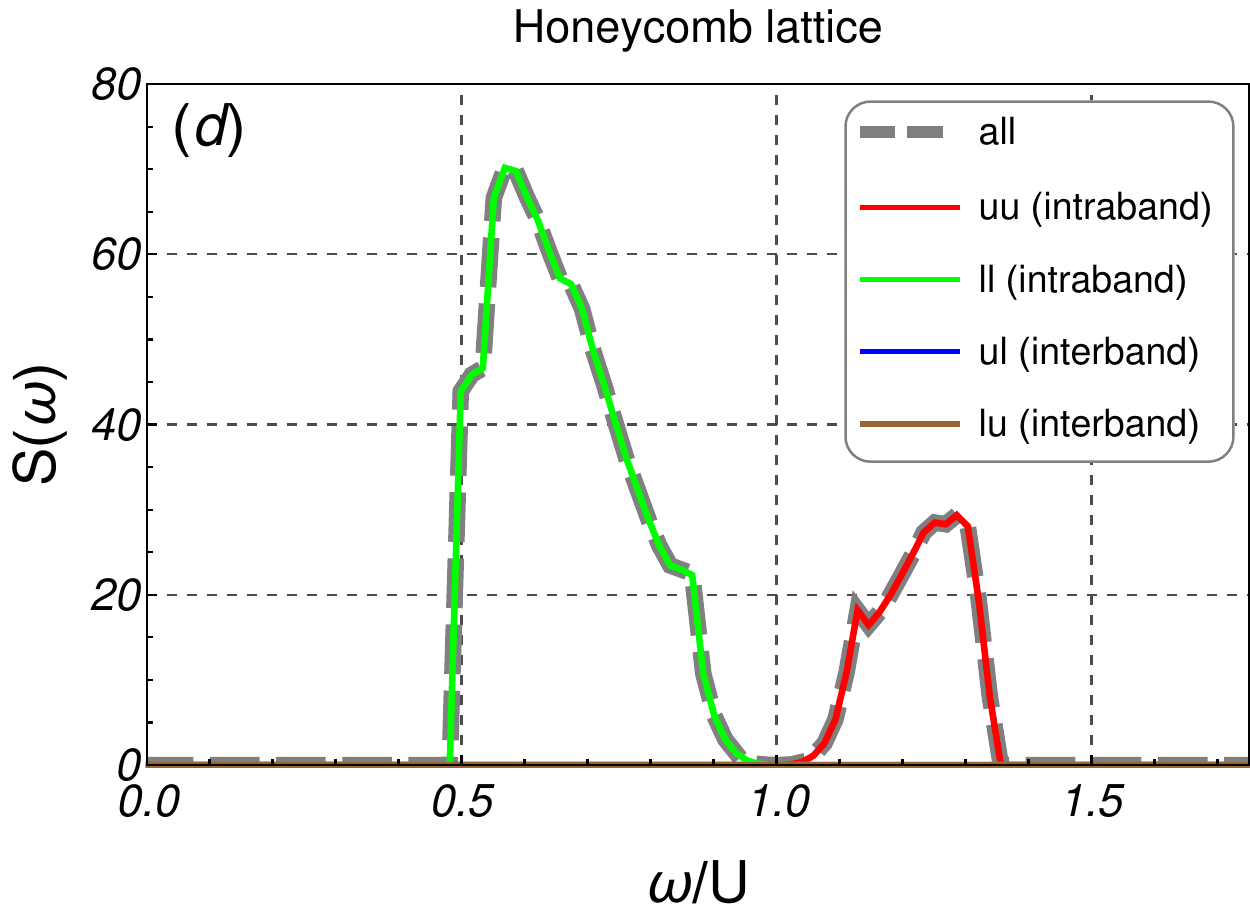}\includegraphics[scale=0.47]{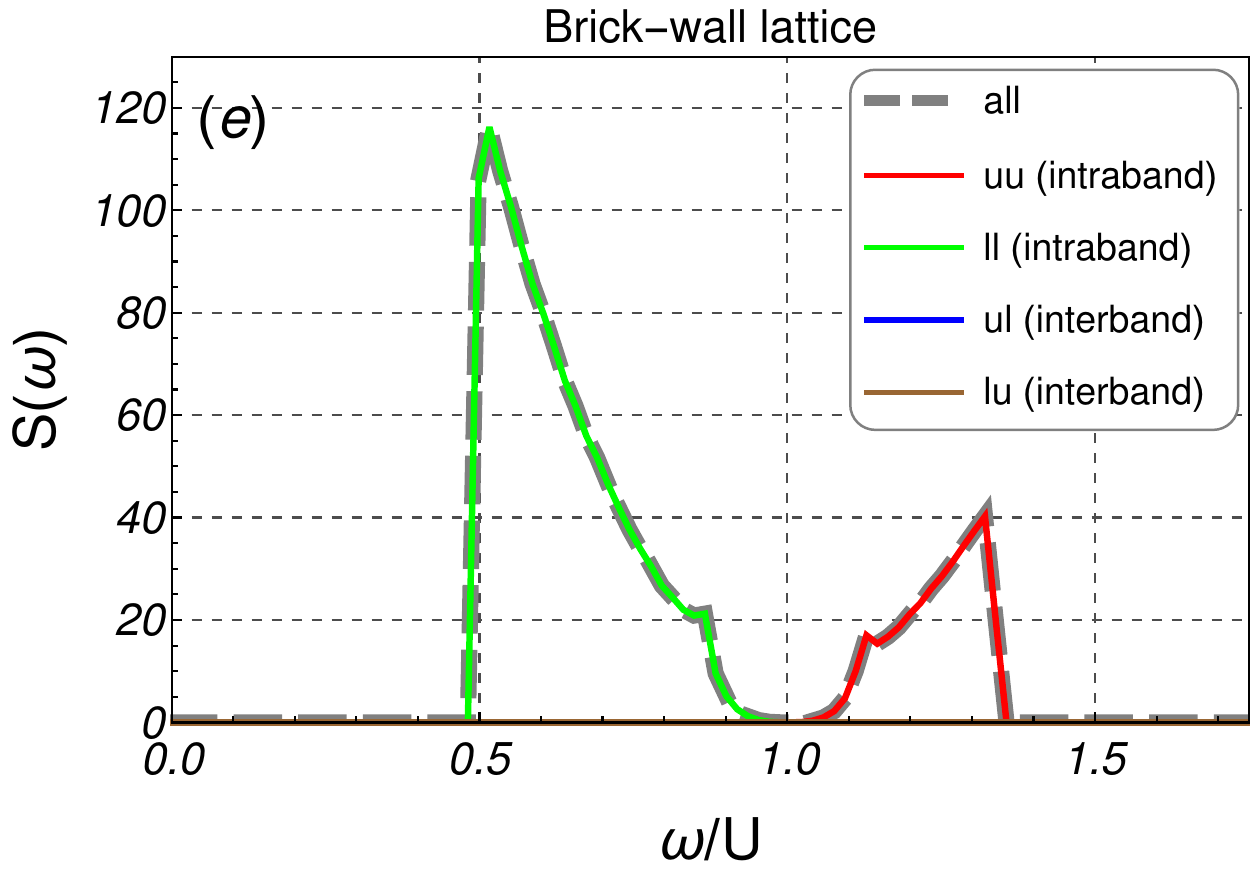}\includegraphics[scale=0.47]{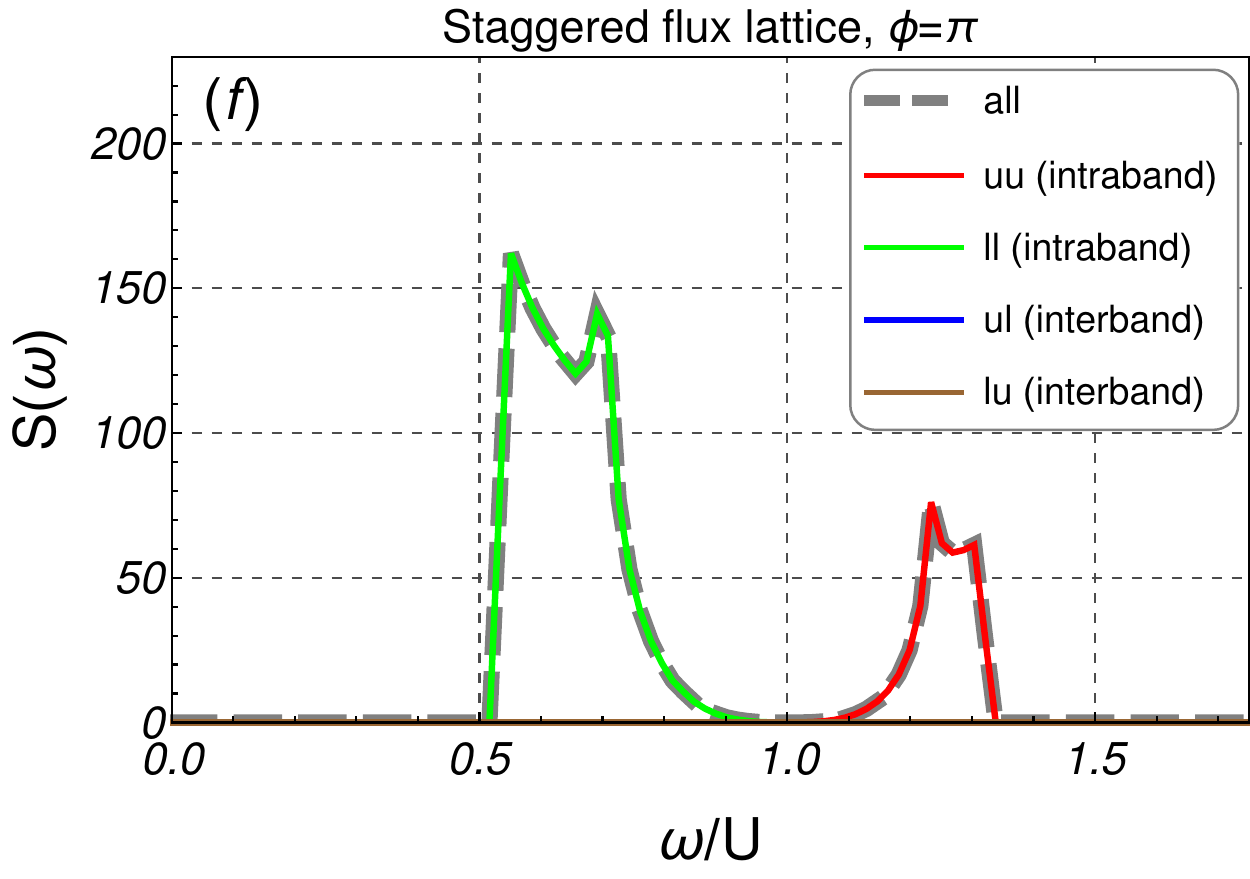}

\caption{(color online) Frequency $\omega$ dependent conductivity $\sigma(\omega)$
(a-c) and $S(\omega)$ (d-f) in the MI phase at zero temperature.
The latter quantity is proportional to the isotropic energy absorption
rate. The remaining parameters are $\mu/U=0.414$, $J/U=0.043$. The
symbols $uu$, $dd$, $ud$, $du$ correspond to the transitions depicted
in Fig. \ref{fig: TRANSITIONS}. The bold and dashed gray line give
the total response of the system coming from the all transitions summed
(i.e. from intra and interband transitions); $\sigma_{q}$ denotes
the quantum unit of conductance defined in Eq. \ref{eq: RE cond}.
\label{fig: COND AND ABSORBTION}}
\end{figure*}

\begin{figure*}[th]
\includegraphics[scale=0.47]{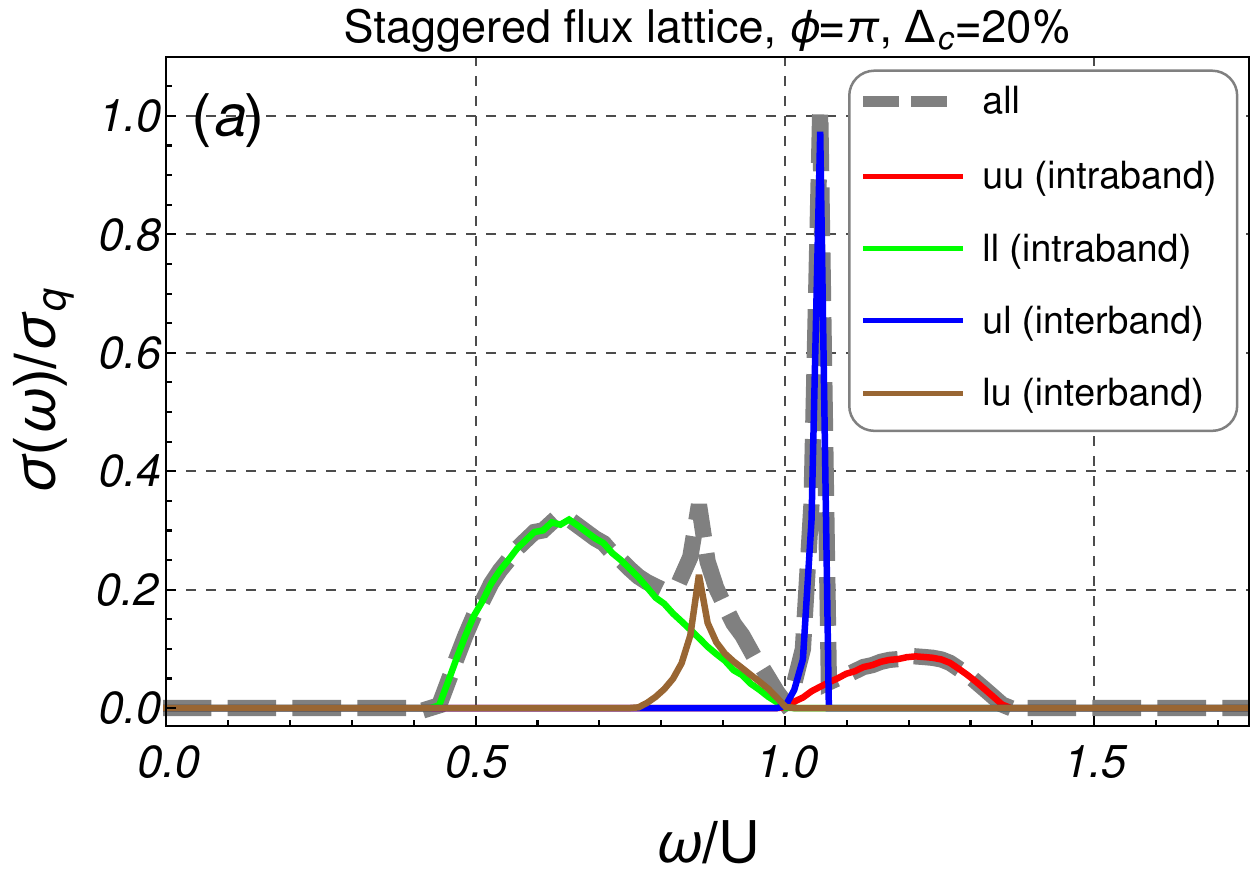}\includegraphics[scale=0.47]{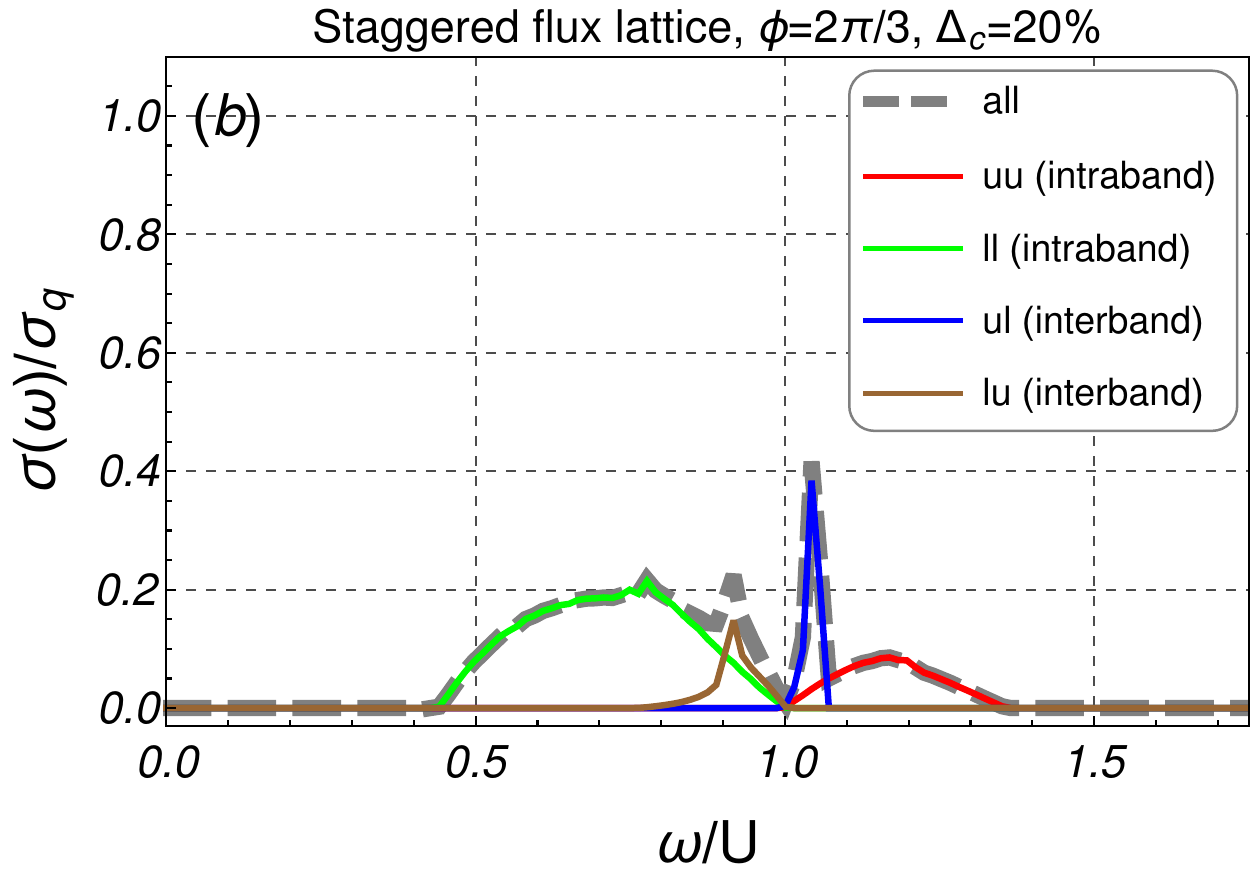}\includegraphics[scale=0.47]{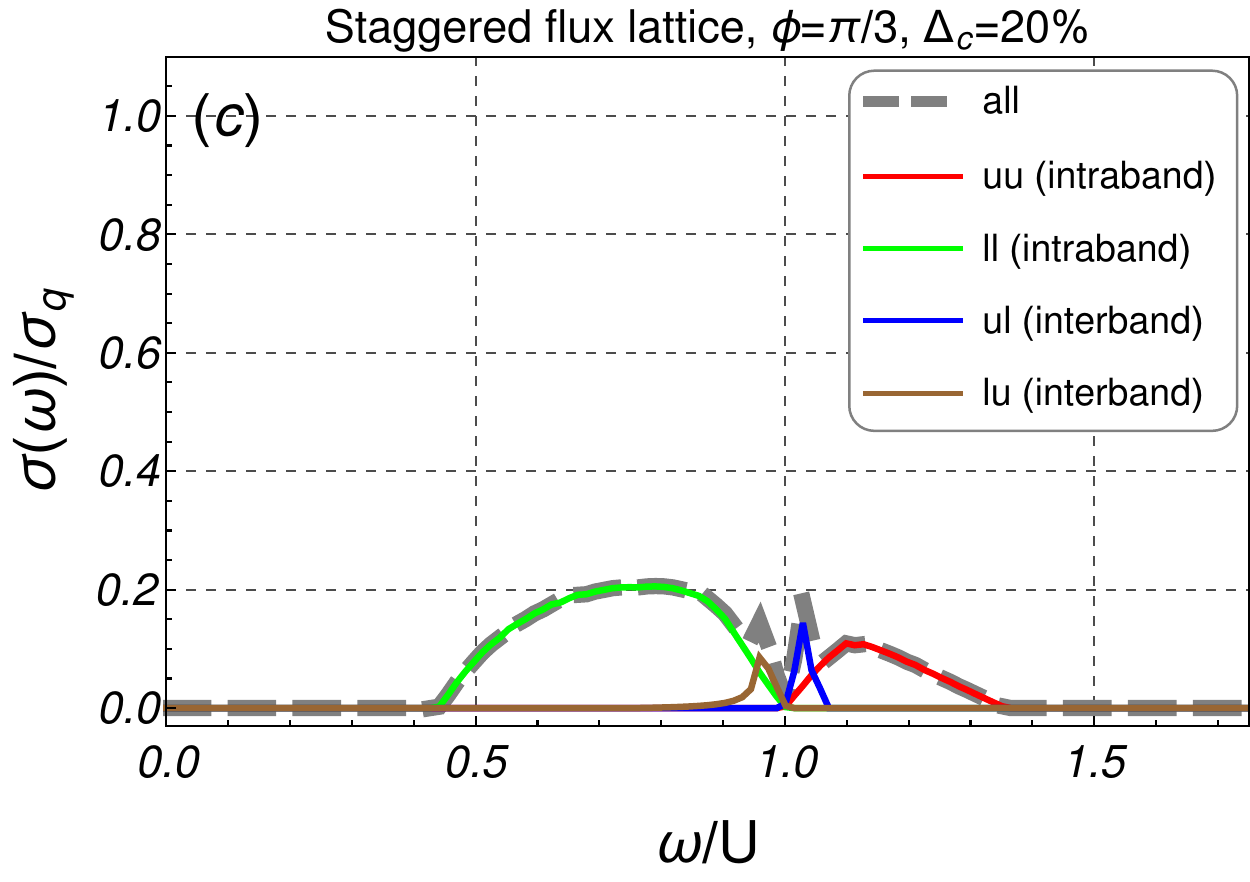}

\caption{(color online) Frequency dependent conductivity in the MI phase at
zero temperature. Staggered flux lattice is considered with the flux
amplitudes: (a) $\phi=\pi$, (b) $\phi=2\pi/3$, (c) $\phi=\pi/3$.
For clarity, we plot dynamical conductivity with the same absolute
detuning of hopping amplitude $J$ from the phase boundary, i.e. $\Delta_{c}=(J_{c}-J)/J_{c}=20\%$
because $J_{c}$ is a $\phi$ dependent function \cite{2008PhRvL.100m0402L,2010PhRvA..81b3404L}.
In particular (a) $J/U\approx0.049$, $(J/U)_{c}\approx0.061$, (b)
$J/U\approx0.040$, $(J/U)_{c}\approx0.050$, (c) $J/U\approx0.036$,
$(J/U)_{c}\approx0.044$. The bold and dashed gray line give the total
response of the system coming from the all transitions summed (i.e.
from intra and interband transitions). The chemical potential is $\mu/U=0.414$.
\label{fig: FLUX TUNING}}
\end{figure*}

\subsection{Linear response kernel in the MI phase \label{sub: Linear-response-kernel}}

The linear response kernel for the phase or amplitude modulation perturbation
(in the MI phase) can be derived from the following form of the four
point correlation function
\begin{eqnarray}
 &  & K\left(i\omega\right)\nonumber \\
 &  & =\frac{1}{N}\sum_{\mathbf{k},\mathbf{k}'}\int_{0}^{\beta}d\tau e^{i\omega\tau}\sum_{\lambda_{1},\lambda_{2},\lambda_{3},\lambda_{4}=1}^{2}\gamma_{\lambda_{1},\lambda_{2}}(\mathbf{k})\gamma_{\lambda_{3},\lambda_{4}}(\mathbf{k}')\nonumber \\
 &  & \times\left\langle \, b_{\lambda_{1},\mathbf{k}}^{*}(\tau)b_{\lambda_{2},\mathbf{k}}(\tau)\, b_{\lambda_{3},\mathbf{k}'}^{*}(0)b_{\lambda_{4},\mathbf{k}'}(0)\right\rangle ,\label{eq: kernel i tau}
\end{eqnarray}
where the average $\left\langle ...\right\rangle $ is defined over
the effective action from Eq.(\ref{eq: effective action}) and we
assume a translation invariant lattice. Applying the Matsubara frequency
transformation for $b$, $b^{*}$ fields, together with the Wick theorem
to Eq. (\ref{eq: kernel i tau}) one gets
\begin{eqnarray}
 &  & K\left(i\omega\right)\nonumber \\
 &  & =\frac{1}{\beta N}\sum_{n}\sum_{\mathbf{k}}\sum_{\lambda_{1},\lambda_{2},\lambda_{3},\lambda_{4}=1}^{2}\gamma_{\lambda_{1},\lambda_{2}}^{1}(\mathbf{k})\gamma_{\lambda_{3},\lambda_{4}}^{2}(\mathbf{k})\nonumber \\
 &  & \times G_{\lambda_{4},\lambda_{1}}^{MI}(\mathbf{k},\omega_{n})G_{\lambda_{2},\lambda_{3}}^{MI}(\mathbf{k},\omega_{n}+\omega),\label{eq: kernel MI}
\end{eqnarray}
in which
\begin{equation}
\left\langle \, b_{\lambda_{\mu},\mathbf{k},n}b_{\lambda_{\nu},\mathbf{k},n}^{*}\right\rangle =G_{\lambda_{\mu},\lambda_{\nu}}^{MI}(\mathbf{k},\omega_{n}),
\end{equation}
where the indices $\lambda_{\mu},\lambda_{\nu}\in\left\{ 1,2\right\} $
denote the matrix element of MI Green function which is the $2$ by
$2$ matrix defined in Eq. (\ref{eq: MI Green funtion in tau}). 

Careful explanation of the form of $K\left(i\omega\right)$ in Eq.
(\ref{eq: kernel i tau}) is needed. Firstly, it describes the paramagnetic
part of linear response in MI phase for which $\omega>0$. Secondly,
the form of $K\left(i\omega\right)$ depends on the form of the considered
external perturbation. In this work, we consider two quantities: $\sigma\left(\omega\right)$
and $S\left(\omega\right)$. The symbol $\sigma\left(\omega\right)$
denotes the dynamical conductivity and is defined by the current autocorrelation
function \cite{1993PhRvB..47..279K}. The function $S\left(\omega\right)$
is proportional to the isotropic energy absorption rate and is defined
by the kinetic energy autocorrelation function \cite{Huber:2007uq,PhysRevLett.109.010401}.
It is important to stress here that $\sigma\left(\omega\right)$ and
$S\left(\omega\right)$ are also related with the response of the
system to the phase and the amplitude periodic modulation of the optical
lattice, respectively \cite{Huber:2007uq,PhysRevLett.109.010401,2011PhRvL.106t5301T}.
Namely, within the notation introduced in Sec. \ref{sub:Staggered-symmetry-lattice },
the optical conductivity can be obtained from
\begin{equation}
\sigma\left(\omega\right)=\left.\frac{e_{eff}^{2}}{\omega}K\left(i\omega\right)\right|_{i\omega\rightarrow\omega+i0^{+}},\label{eq: conductivity - general form}
\end{equation}
where
\begin{equation}
\gamma_{\lambda_{1},\lambda_{2}}^{1}(\mathbf{k})=\gamma_{\lambda_{1},\lambda_{2}}^{2}(\mathbf{k})=\partial_{k_{x}}f{}_{\lambda_{1},\lambda_{2}}(\mathbf{k}),\label{eq: gamma cond 1}
\end{equation}
and $e_{eff}$ is the effective charge (which in the optical lattice
can be generated e.g. by the synthetic gauge field). The expression
$i\omega\rightarrow\omega+i0^{+}$ denotes the analytic continuation
and $f{}_{\lambda_{1},\lambda_{2}}(\mathbf{k})$ is the $(\lambda_{1},\lambda_{2})$
matrix element of $\mathbf{F}(\mathbf{k})$ from Eq. (\ref{eq: general TB}).
Partial derivative over $k_{x}$ in Eq. (\ref{eq: gamma cond 1})
means that $xx$ component of the longitudinal conductivity is considered.
The form of Eq. (\ref{eq: conductivity - general form}) agrees with
those ound in literature \cite{1993PhRvB..47..279K,2014PhRvA..89b3631S,PhysRevB.96.094520}.

In the case of $S(\omega)$ one gets
\begin{equation}
S\left(\omega\right)=\textrm{Im}\left[\left.K\left(i\omega\right)\right|_{i\omega\rightarrow\omega+i0^{+}}\right],\label{eq: absorption energy}
\end{equation}
where
\begin{equation}
\gamma_{\lambda_{1},\lambda_{2}}^{1}(\mathbf{k})=\gamma_{\lambda_{1},\lambda_{2}}^{2}(\mathbf{k})=f{}_{\lambda_{1},\lambda_{2}}(\mathbf{k}).\label{eq: gamma absorption 1}
\end{equation}
$S\left(\omega\right)$ has been previously studied in different contexts,
e.g. correlated fermions \cite{PhysRevA.74.041604}, Higgs mode \cite{PhysRevLett.109.010401}
or thermometry in MI phase \cite{PhysRevA.94.043612}.

Because we are focused on the MI phase, one can give the general form
of $K\left(i\omega\right)$ by using Eqs. (\ref{eq: kernel MI}) and
(\ref{eq: Green function diagonal})-(\ref{eq:gap}), i.e.
\begin{eqnarray}
 &  & K\left(i\omega\right)\label{eq: kernel general form iomega}\\
 &  & =\frac{1}{N}\sum_{\mathbf{k}}\sum_{\alpha,\beta\in\{u,l\}}\Gamma_{\alpha\beta}^{\gamma}\left(\mathbf{k}\right)\frac{\left[1-z_{p,\alpha}\left(\mathbf{k}\right)\right]z_{p,\beta}\left(\mathbf{k}\right)}{E_{h,\alpha}\left(\mathbf{k}\right)+i\hbar\omega-E_{p,\beta}\left(\mathbf{k}\right)},
\end{eqnarray}
where
\begin{eqnarray}
 &  & \Gamma_{uu}^{\gamma}\left(\mathbf{k}\right)=\Gamma_{ll}^{\gamma}\left(\mathbf{k}\right)=\nonumber \\
 &  & =\left(\frac{\gamma_{2,1}^{1}\left(\mathbf{k}\right)f{}_{1,2}\left(\mathbf{k}\right)+\gamma_{1,2}^{1}\left(\mathbf{k}\right)\bar{f}_{1,2}\left(\mathbf{k}\right)}{2\left|f_{1,2}\left(\mathbf{k}\right)\right|}\right)^{2},\label{eq: coeficient uu}
\end{eqnarray}
\begin{eqnarray}
 &  & \Gamma_{ul}^{\gamma}\left(\mathbf{k}\right)=\Gamma_{lu}^{\gamma}\left(\mathbf{k}\right)\nonumber \\
 &  & =-\left(\frac{\gamma_{1}^{2,1}\left(\mathbf{k}\right)f{}_{1,2}\left(\mathbf{k}\right)-\gamma_{1}^{1,2}\left(\mathbf{k}\right)\bar{f}_{1,2}\left(\mathbf{k}\right)}{2\left|f_{1,2}\left(\mathbf{k}\right)\right|}\right)^{2},\label{eq: kernel general form iomega end}
\end{eqnarray}
and weights $\Gamma_{uu}^{\gamma}$, $\Gamma_{ll}^{\gamma}$ and $\Gamma_{ul}^{\gamma}$,
$\Gamma_{lu}^{\gamma}$ correspond to the intra and inter-band transitions
depicted in Fig. (\ref{fig: TRANSITIONS}) for which a more detailed
discussion will be given shortly (the $\gamma$ index is reserved
for conductivity $\Gamma_{\alpha\beta}^{cond}$ and for energy absorption
rate $\Gamma_{\alpha\beta}^{abs}$). Moreover, in derivation of Eqs.
(\ref{eq: kernel general form iomega})-(\ref{eq: kernel general form iomega end})
we assume a zero temperature limit and for simplicity we identify
$\gamma^{1}$ with $\gamma^{2}$ i.e. $\gamma^{1}=\gamma^{2}$ which
agrees with Eqs. (\ref{eq: gamma cond 1}) and (\ref{eq: gamma absorption 1}).

It is also worth pointing out here that Eqs. (\ref{eq: kernel general form iomega}-\ref{eq: kernel general form iomega end})
are valid for an arbitrary form of the staggered symmetry introduced
by the hopping amplitude $\mathbf{F}(\mathbf{k})$ (see Eq. (\ref{eq: general TB}))
and therefore, these results are not restricted to the lattices considered
here (i.e. for the lattices defined by Eqs. (\ref{eq: lattice flux})-(\ref{eq: lattice brick-wall})).

\subsection{Dynamical conductivity \label{sub:Dynamical-conductivity}}

To calculate the dynamical (longitudinal) conductivity $\sigma(\omega)$
from Eq. (\ref{eq: conductivity - general form}), the analytic continuation
of $K\left(i\omega\right)/\omega$ is needed ($i\omega\rightarrow\omega+i0^{+}$).
This yields the following form of $xx$ component of the longitudinal
conductivity
\begin{eqnarray}
 &  & \textrm{Re}\sigma\left(\omega\right)\nonumber \\
 &  & =\sigma_{q}\frac{2\pi^{2}}{N}\sum_{\mathbf{k}}\sum_{\alpha,\beta\in\{u,l\}}\Gamma_{\alpha\beta}^{cond}\left(\mathbf{k}\right)\frac{\left[1-z_{p,\alpha}\left(\mathbf{k}\right)\right]z_{p,\beta}\left(\mathbf{k}\right)}{E_{h,\alpha}\left(\mathbf{k}\right)-E_{p,\beta}\left(\mathbf{k}\right)}\nonumber \\
 &  & \times\delta\left(\omega-\left[E_{p,\beta}\left(\mathbf{k}\right)-E_{h,\alpha}\left(\mathbf{k}\right)\right]\right),\label{eq: RE cond}
\end{eqnarray}
where we focus on the real part of dynamical conductivity, i.e. $\textrm{Re}\sigma\left(\omega\right)$
and we define $\sigma_{q}=e_{eff}^{2}/h$ as a quantum unit of conductance
\cite{2014PhRvA..89b3631S}. From the form of $\textrm{Re}\sigma\left(\omega\right)$,
we see that the response of the system appears at the energy difference
between quasiparticle ($E_{p,\beta}\left(\mathbf{k}\right)$) and
hole excitations ($E_{h,\alpha}\left(\mathbf{k}\right)$) through
the Dirac delta function $\delta$. Therefore for two band models
in the MI phase, one can have four types of excitations. The transitions
corresponding to these excitations are schematically drawn in Fig.
(\ref{fig: TRANSITIONS}). In general, one can divide such transitions
into two classes which have intra ($uu$, $ll$) or interband ($ul$,
$lu$) character. The intraband transitions are mostly responsible
for the lowest and highest energy excitations and e.g. they can describe
remarkable phenomenon like the finite critical conductivity \cite{PhysRevLett.64.587,1993PhRvB..47..279K,PhysRevB.49.9794,2014PhRvA..89b3631S}.
The interband transitions are responsible for intermediate behavior
and in the cases considered in this work, they strongly modify the
response around the energy scales in which the Dirac points appear
(a similar behavior has been very recently reported in the uniform
magnetic fields \cite{PhysRevB.96.094520} when this work was finalized).

Next, calculating the conductivity coefficients $\Gamma_{\alpha\beta}^{cond}$
in Eqs. (\ref{eq: RE cond}) by using Eqs. (\ref{eq: coeficient uu})-(\ref{eq: kernel general form iomega end})
and (\ref{eq: gamma cond 1}) one gets
\begin{equation}
\Gamma_{uu}^{cond}\left(\mathbf{k}\right)=\Gamma_{ll}^{cond}\left(\mathbf{k}\right)=\left(\partial_{k_{x}}\left|f_{1,2}\left(\mathbf{k}\right)\right|\right)^{2},\label{eq: gamma uu ll cond}
\end{equation}
\begin{eqnarray}
 &  & \Gamma_{ul}^{cond}\left(\mathbf{k}\right)=\Gamma_{lu}^{cond}\left(\mathbf{k}\right)=\nonumber \\
 &  & =-\left(\frac{f{}_{1,2}(\mathbf{k})\partial_{k_{x}}\bar{f}{}_{1,2}(\mathbf{k})-\bar{f}_{1,2}(\mathbf{k})\partial_{k_{x}}f{}_{1,2}(\mathbf{k})}{2\left|f_{1,2}\left(\mathbf{k}\right)\right|}\right)^{2}.\label{eq: gamma ul lu cond}
\end{eqnarray}
From the above equations, we see that all four transitions ($uu$,
$ll$, $ul$, $lu$) can contribute to the conductivity. However,
to consider a particular lattice, explicit forms of the functions
$\Gamma_{\alpha\beta}^{cond}$  have to be given together with the
remaining functions in Eq. (\ref{eq: RE cond}).

To show how the  intra ($uu$, $ll$) and interband ($ul$, $lu$)
transitions behave in the lattices with staggered symmetry, we consider
at first the honeycomb type of lattices (see Eqs. (\ref{eq: lattice honeycomb})
and (\ref{eq: lattice brick-wall})). For standard honeycomb lattice
one gets
\begin{eqnarray}
 &  & \Gamma_{uu}^{cond}\left(\mathbf{k}\right)=\Gamma_{ll}^{cond}\left(\mathbf{k}\right)=3J^{4}\sin^{2}\left(\frac{\sqrt{3}}{2}k_{x}\right)\nonumber \\
 &  & \times\left(\frac{2\cos\left(\frac{\sqrt{3}}{2}k_{x}\right)+\cos\left(\frac{3}{2}k_{y}\right)}{\left|f_{1,2}^{hc}\left(\mathbf{k}\right)\right|}\right)^{2},
\end{eqnarray}
\begin{equation}
\Gamma_{ul}^{cond}\left(\mathbf{k}\right)=\Gamma_{lu}^{cond}\left(\mathbf{k}\right)=\frac{3J^{4}\sin^{2}\left(\frac{\sqrt{3}}{2}k_{x}\right)\sin^{2}\left(\frac{3}{2}k_{y}\right)}{\left|f_{1,2}^{hc}\left(\mathbf{k}\right)\right|^{2}},
\end{equation}
and for brick-wall lattice
\begin{eqnarray}
 &  & \Gamma_{uu}^{cond}\left(\mathbf{k}\right)=\Gamma_{ll}^{cond}\left(\mathbf{k}\right)=4J^{4}\sin^{2}\left(k_{x}\right)\nonumber \\
 &  & \times\left(\frac{2\cos\left(k_{x}\right)+\cos\left(k_{y}\right)}{\left|f_{1,2}^{bw}\left(\mathbf{k}\right)\right|}\right)^{2},
\end{eqnarray}
\begin{equation}
\Gamma_{ul}^{cond}\left(\mathbf{k}\right)=\Gamma_{lu}^{cond}\left(\mathbf{k}\right)=\frac{4J^{4}\sin^{2}\left(k_{x}\right)\sin^{2}\left(k_{y}\right)}{\left|f_{1,2}^{bw}\left(\mathbf{k}\right)\right|^{2}},
\end{equation}
where we set the lattice constant $a$ to one. It is straightforward
to see that the above formulas for honeycomb and brick-wall lattices,
have a similar form (see also Eqs. (\ref{eq: lattice honeycomb})
and (\ref{eq: lattice brick-wall})) which result in similar behavior
of dynamical conductivity - Figs. \ref{fig: COND AND ABSORBTION}
a and b. In these Figures one sees that interband transitions ($ul$,
$lu$), which appear near the Dirac point, are more pronounced than
the intraband transitions ($uu$, $ll$) (Dirac point is visible as
a vanishing of conductivity at $\omega=U$). Similar behavior of significant
differences in the intra and interband transition amplitudes has been
very recently reported also for the uniform magnetic field \cite{PhysRevB.96.094520}.
Moreover, the shifting of Dirac points in the wave vector space $\mathbf{k}$
(Fig. \ref{fig: tightbinding and diracs} a), slightly changed the
interband transitions amplitude, i.e. the brick-wall interband transitions
had slightly higher amplitudes than the corresponding transitions
for the honeycomb lattice. Therefore, the modification of the lattice
geometry had a direct consequence on the linear response dynamics
near the Dirac points.

We can achieve much more pronounced tuning of the dynamics near the
Dirac points if we consider the staggered flux lattice. Prior to showing
this, let us first write the coefficients $\Gamma_{\alpha\beta}^{cond}$
for this type of lattice:
\begin{eqnarray}
 &  & \Gamma_{uu}^{cond}\left(\mathbf{k}\right)=\Gamma_{ll}^{cond}\left(\mathbf{k}\right)=16J^{4}\sin^{2}\left(k_{x}a\right)\nonumber \\
 &  & \times\left(\frac{\cos\left(k_{x}a\right)+\cos\left(\phi/2\right)\cos\left(k_{y}a\right)}{\left|f_{1,2}^{flux}\left(\mathbf{k}\right)\right|}\right)^{2},
\end{eqnarray}
\begin{eqnarray}
 &  & \Gamma_{ul}^{cond}\left(\mathbf{k}\right)=\Gamma_{lu}^{cond}\left(\mathbf{k}\right)=\nonumber \\
 &  & =\frac{4J^{4}\sin^{2}\left(k_{x}a\right)\cos^{2}\left(k_{y}a\right)\sin^{2}\left(\phi/2\right)}{\left|f_{1,2}^{flux}\left(\mathbf{k}\right)\right|^{2}}.
\end{eqnarray}
From the above equations, we see that the interband coefficients,
i.e. $\Gamma_{ul}^{cond}\left(\mathbf{k}\right)$ and $\Gamma_{lu}^{cond}\left(\mathbf{k}\right)$,
depend on the square power of the $\sin\left(\phi/2\right)$ function.
This explains why the interband transitions are the largest for $\phi=\pi$
and they vanish in the limit $\phi\rightarrow0$ (for $\phi=0$ the
square lattice limit is recovered where interband transitions do not
exist \cite{2014PhRvA..89b3631S}). We prove this remarkable behavior
of interband transitions by plotting the dynamical conductivity for
different values of $\phi$ amplitude in Fig. \ref{fig: FLUX TUNING}.
It is important to point out here that such a behavior is directly
connected to the steepness of tight-binding dispersion near the Dirac
points which is tuned with a different $\phi$ amplitude, see Fig.
\ref{fig: tightbinding and diracs} b.

Moreover, it is very instructive to explain why the interband transitions
($ul$, $lu$) appear around the Dirac points (i.e. around $\omega=U$)
(see Figs. \ref{fig: COND AND ABSORBTION} a-b). If we look at Fig.
\ref{fig: TRANSITIONS}, we see that the bandwidth of qasiparticle
excitations is by about twice larger than the quasihole band. This
can be simply accounted for by the different effective hoppings of
particles and holes and e.g. for free bosonic case it is simply $(n_{0}+1)J$
and $n_{0}J$, respectively (see, e.g., \cite{PhysRevA.93.013622,PhysRevLett.100.216401}).
Therefore, moving away from the Dirac points which for staggered flux
lattice are at $(\pm\pi/2,\pm\pi/2)$ (see also Fig. \ref{fig: TRANSITIONS}),
we see that the $ul$ and $lu$ transitions for a given $\mathbf{k}$
vary slightly from the interaction energy $U$ because of the different
effective hoppings of particles and holes. We conclude, that exactly
this difference in the effective hoppings, gives rise to the linear
response dynamics which concentrates around the Dirac points. A corresponding
situation can be also found for the honeycomb type of lattices.

\subsection{Isotropic energy absorption rate \label{sub: Isotropic-energy-absorption}}

In Sec. \ref{sub:Dynamical-conductivity} we have shown how the interband
transitions can be engineered by the lattice geometry or staggered
flux. Here, we show that such a manipulation of transitions in the
strongly correlated bosonic system can be also achieved in a much
robust way, i.e. by a suitable choice of external perturbation. Dynamical
conductivity considered in the previous section is related to the
periodic phase modulation of the optical lattice \cite{2011PhRvL.106t5301T}.
Now we show what happens if we consider the periodic amplitude lattice
modulation which is directly connected to the function $S(\omega)$
(Eq. (\ref{eq: absorption energy})) \cite{PhysRevA.74.041604,Huber:2007uq,PhysRevLett.109.010401}.

Namely, we consider $S(\omega)$ which is proportional to the isotropic
energy absorption rate and is defined in Eqs. (\ref{eq: absorption energy})-(\ref{eq: kernel general form iomega end}).
After analytical calculations one gets
\begin{eqnarray}
 &  & S\left(\omega\right)\nonumber \\
 &  & =\frac{\pi}{N}\sum_{\mathbf{k}}\sum_{\alpha,\beta\in\{u,l\}}\Gamma_{\alpha\beta}^{cond}\left(\mathbf{k}\right)\left[1-z_{p,\alpha}\left(\mathbf{k}\right)\right]z_{p,\beta}\left(\mathbf{k}\right)\nonumber \\
 &  & \times\delta\left(\omega-\left[E_{p,\beta}\left(\mathbf{k}\right)-E_{h,\alpha}\left(\mathbf{k}\right)\right]\right),\label{eq: Im absorption}
\end{eqnarray}
where
\begin{equation}
\Gamma_{uu}^{abs}\left(\mathbf{k}\right)=\Gamma_{ll}^{abs}\left(\mathbf{k}\right)=\left(\left|f_{1,2}\left(\mathbf{k}\right)\right|\right)^{2},\label{eq: gamma uu ll absorb}
\end{equation}
\begin{eqnarray}
 &  & \Gamma_{ul}^{abs}\left(\mathbf{k}\right)=\Gamma_{lu}^{abs}\left(\mathbf{k}\right)=0,\label{eq: gamma ul lu absorb}
\end{eqnarray}
and where the $\left|f_{1,2}\left(\mathbf{k}\right)\right|$ function
is the tight-binding energy dispersion of the lattices considered
in this work (see, Eqs (\ref{eq: hc - tight-binding})-(\ref{eq: flux - tight-binding})).
The result of Eq. (\ref{eq: gamma ul lu absorb}) is especially interesting.
It says that interband transtions ($ul$, $lu$) are completely turned
off in the isotropic and periodic modulation of the lattice amplitude
for the lattices with staggered symmetry introduced by the hopping
amplitude (see Eq. (\ref{eq: general TB}). We confirm this analytical
result, by plotting Eqs. (\ref{eq: Im absorption})-(\ref{eq: gamma ul lu absorb})
in Figs. \ref{fig: COND AND ABSORBTION} d, e and f for the honeycomb,
brick-wall and staggered flux lattice, respectively. These plots should
be compared with the conductivity $\sigma(\omega)$ plots, Figs. \ref{fig: COND AND ABSORBTION}
a, b and c, for which the interband transitions are present (for $\sigma(\omega)$
the interband weights $\Gamma_{ul}^{cond}$ and $\Gamma_{lu}^{cond}$
do not vanish identically like for the $S(\omega)$ case, compare
Eqs. (\ref{eq: gamma ul lu cond}) and (\ref{eq: gamma ul lu absorb})).

\section{Summary \label{sec: Summary}}

In this work we have considered linear response dynamics in the MI
phase for the lattices with staggered symmetry. This symmetry was
introduced by the hopping amplitude and takes into account a broad
class of lattice currently realized in the optical lattice systems.
We have focused on the lattices which contain Dirac points in their
tight-binding energy dispersions and which can allow Dirac like physics
engineering.

As the main result of our considerations we have shown that linear
response dynamics around the Dirac points can be highly tuned by a
suitable type of the external perturbation or by modification of the
lattice parameters. This shows that the interband transitions can
be very sensitive to the particular experiment realization and can
be a signature for emergence of relativistic dynamics related with
the Dirac points.

Moreover, we have explained that such a peculiar dynamics around the
Dirac points is directly connected to the effective hopping energies
for holes and particles. This further can provide an indirect proof
of the non-equal values of bandwidth for quasiparticle and hole excitations
in the MI phase.

It is also important to stress here that presented theoretical framework
can be directly applied to the other lattice systems showing the staggered
lattice symmetry.
\begin{acknowledgments}
We are grateful to R. Micnas, T. P. Polak for valuable discussions
and carefully reading of the manuscript. This work was supported by
the National Science Centre, Poland, project no. 2014/15/N/ST2/03459.
\end{acknowledgments}

\section{Appendix}

\subsection{Tight-binding energy dispersions \label{sub:Tight-binding-energy-dispersions}}

\begin{widetext}Tight-binding energy dispersions which are plotted
in Fig. \ref{fig: tightbinding and diracs} have the following forms:

- for honeycomb lattice

\begin{equation}
\left|f_{1,2}^{hc}\left(\mathbf{k}\right)\right|=\pm J\sqrt{1+4\cos^{2}\left(\frac{\sqrt{3}}{2}k_{x}a\right)+4\cos\left(\frac{\sqrt{3}}{2}k_{x}a\right)\cos\left(\frac{3}{2}k_{y}a\right)},\label{eq: hc - tight-binding}
\end{equation}

- for brick-wall lattice
\begin{equation}
\left|f_{1,2}^{bw}\left(\mathbf{k}\right)\right|=\pm J\sqrt{1+4\cos^{2}\left(k_{x}a\right)+4\cos\left(k_{x}a\right)\cos\left(k_{y}a\right)},
\end{equation}

- for staggered flux lattice
\begin{equation}
\left|f_{1,2}^{flux}\left(\mathbf{k}\right)\right|=\pm2J\sqrt{\cos^{2}\left(k_{x}a\right)+\cos^{2}\left(k_{y}a\right)+2\cos\left(\frac{\phi}{2}\right)\cos\left(k_{x}a\right)\cos\left(k_{y}a\right)},\label{eq: flux - tight-binding}
\end{equation}
where $a$ is the lattice constant.\end{widetext}

\bibliographystyle{apsrev4-1}
\bibliography{library}

\end{document}